\documentclass[11pt,a4paper,preprint]{revtex4}
\usepackage{amsmath}
\usepackage{amsfonts}
\usepackage{amssymb}
\usepackage{graphicx}
\usepackage{epstopdf}
\usepackage{xcolor}
\normalsize
\usepackage{siunitx, booktabs}
\graphicspath{{./figures}}
\usepackage{hyperref}
\usepackage{tikz}
\usepackage{ulem}
\usetikzlibrary{decorations.pathreplacing}
\usepackage[margin=10pt,font=small]{caption}

\usepackage[left=2cm,right=2cm,top=2cm,bottom=2cm]{geometry}
\renewcommand{\u}{\mathbf{u}}
\renewcommand{\v}{\mathbf{v}}
\newcommand{\into}{\int_{\Omega}}
\newcommand{\intoi}{\int_{\Omega_i}}
\DeclareMathOperator{\Tr}{tr}

\begin{document}
	
	\title{The influence of finite size particles on fluid velocity and transport though porous media}
	
	\author{M. Residori}
	\author{S. Praetorius}
	\affiliation{Institute of Scientific Computing, TU Dresden, 01062 Dresden, Germany}
	
	\author{P. de Anna}
	\affiliation{Institute of Earth Sciences, University of Lausanne, Lausanne 1015, Switzerland}
	
	\author{A. Voigt}
	\affiliation{Institute of Scientific Computing, TU Dresden, 01062 Dresden, Germany and Dresden Center for Computational Materials Science (DCMS), TU Dresden, 01062 Dresden, Germany}
	
	\begin{abstract}
		Understanding the coupling between flow, hydrodynamic transport and dispersion of colloids of finite size in porous media is a long-standing challenge. This problem is relevant for a broad range of natural and engineered processes, including contaminant and colloidal transport, mixing of bio-chemical compounds, kinetics of  reactions and groundwater bio-remediation, but also transport phenomena related to different systems like membranes, or blood flow. While classical models for colloidal transport rely on macro-dispersion theory and do not take into consideration the complex and heterogeneous structure of the porous host medium, recent studies take into consideration the detailed structure of the porous system and its impact on fluid velocity. However, the impact of confinement conditions, represented by the ratio of the radius of particles $a$ and pore throat size $\lambda$, has been overlooked. Here, we use numerical simulations of fluid particle dynamics in resolved porous media to demonstrate that particle confinement affects the fluid macroscopic velocity field which in turn affects the particle transport itself. Our results show that even under small confinement conditions ($a/\lambda \sim 2$~\%), fluid and transported particles are dynamically re-routed towards more permeable paths. This leads to the emergence of ephemeral laminar vortexes at pore throat entrances and affects the variance and mean fluid velocity.
	\end{abstract}
	
	\maketitle
	
	\section{Introduction}
	\label{intro}
	
	Most engineered and natural systems characterized by a porous structure can host fluids that, driven by a macroscopic pressure gradient, move through their network of pores~\cite{beardynamics1988}. In most scenarios, the concerned fluid is water that, being an excellent solvent, is carrying dissolved substances or suspended particles, whose transport rate controls several phenomena, including mixing, reaction kinetics and filtration~\cite{Molnar2005}. The characteristics of these phenomena are a direct consequence of the underlying pore structure. A common issue of such pore structures is their inherent structural heterogeneity~\citep{Dentz2011}, or spatial variability, that has been shown to emerge from sub-micron to meter scales~\citep{freezecherry}. A wide range of variability appears in the distribution of individual pores size (i.e. the space among solid grains, available for fluid motion). 
	
	Since within such porous systems fluid velocities are typically low (on the order of \SI{1}{\metre\per\day}, equivalent to about \SI{20}{\micro\metre\per\second}) the flow of water can be assumed to be laminar on average length scale, controlled by viscous forces rather than inertia, and stationary~\citep{beardynamics1988}. Under such conditions, an analytical relationship between the pore throat size distribution $f_\lambda \sim \lambda^{-\beta}$ and the distribution of low fluid velocities $f_u \sim u^{-\beta/2}$ has been suggested in \citep{deAnnaPRF2017}. Under the assumption of Stokes flow it has been shown that each pore-throat hosting a net transfer of fluid has the same statistical distribution of velocities as the one of a pipe, called the porelet, so that the overall probability density function (PDF) of fluid velocities within the heterogeneous structure can be determined by the superposition of all porelets in two-dimensional~\citep{deAnnaPRF2017} and three-dimensional~\citep{DentzJFM2018} systems. This relationship, which allows us to make predictions within a continuous-time random-walk framework for the asymptotic statistics of the spreading of fluid particles along their own trajectories, is based on the linearity of the Stokes equation and assumes dissolved substances and suspended particles to have no impact on the stationary flow field. The assumption at least becomes questionable if instead of average length scale the spatial variable pore throat size and the impact of the particles on the flow are considered.
	
	The spatial variability of fluid velocity, which measures the overall velocity contrast between pores or between channels of high velocity and zones of stagnation, has a major control on the transport of substances, dissolved or suspended~\citep{Dentz2011}. While classical macro dispersion theories have been shown to adequately represent relatively homogeneous media (e.g. the ones characterized by a well defined scale for average pore size)~\cite{beardynamics1988}, it is known that such model predictions diverge from the observation of transport through complex porous systems that results in anomalous transport properties as early arrival times and long tailing~\cite{Berkowitz2006,Dentz2011}. In such media the observed transport also impacts mixing kinetics~\cite{deAnnaEST2014,Heyman2020}, and mixing-driven processes as filtration~\cite{Nishiyama2012,miele2019}, or microbial dispersal~\citep{Scheidweiler2020,deAnnaNaturePhys2021}. In several scenarios, flowing water carries suspended colloidal particles and microbial cells or aggregates. The size of such suspended particles and cell aggregates can vary over several orders of magnitude, affecting their sedimentation~\citep{morrisbook,bergbook} and overall transport~\citep{Sirivithayapakorn2003,Auset2004}. In particular, enhanced suspended particle velocity in confined environments has been observed and associated with their finite-size, also known as size exclusion effect~\citep{Prieve1978,Babakhani2019}. This phenomenon forces larger particles to remain within channels of high flow while preventing them from accessing pores of low velocity whose size is comparable to the particle diameter. The macroscopic effect is to increase the average suspended particle velocity, which modifies the macroscopic breakthrough curves~\citep{Keller2004} and dispersion coefficient~\citep{James2003}. These phenomena clearly breakdown the assumption that transported particles do not affect the stationary velocity field, as in~\citep{deAnnaPRF2017,miele2019,Scheidweiler2020} that relate the medium physical structure to the fluid and suspended particles velocity to predict macroscopic transport. 
	
	Here, we investigate the impact of transported finite-size particles on fluid velocity distribution and dynamics, by means of a novel fluid particle numerical scheme for simulating flow and transport through complex porous structures that takes particle-particle and particle-fluid interactions into account. The model includes inertial effects as well, although they are generally negligible in the regime considered in this work. We consider the local feedback of particles with a finite radius $a$, smaller but comparable to the local pore throat size $\lambda$, on the fluid velocity distribution. The no-slip boundary conditions at grain walls exert a viscous drag on finite-size particles when facing a constriction (pore opening) of comparable size. Thus, a dynamic interaction between the passage of finite-size particles through such pores and the flow field is expected. We point out that the model employed in this work simulates soft particles whose dynamic can be regarded as solid particle dynamic as discussed in Sec.~\ref{method}. Our modeling does not include clogging or jamming effects as throat sizes are comparable, but larger than the particle radius. We numerically investigate the dynamical change of fluid and particle velocity in laminar conditions. We compare our simulation results with Stokes flow and point particles. We show that for even small confinement conditions ($a/\lambda \sim 2$~\%), fluid and transported particles are dynamically re-routed towards more permeable paths. This leads to the emergence of ephemeral laminar vortexes at pore throat entrances and affects the variance and mean fluid velocity. We set the physical problem and present the results in terms of adimensional quantities, as detailed in the method section.
	
	\section{Methods} \label{sec:2}
	Natural and engineered porous systems have a three-dimensional structure that is often heterogeneous, i.e. spatially variable. The latter is typically characterized by grains of different sizes that are randomly packed. As a consequence, the pore space among grains is also heterogeneously distributed. Mostly for practical reasons (experimental and/or numerical), studies on porous media flow and flow-driven processes are often conducted in two-dimensional replicates that mimic key structural features, such as the pore size distribution~\cite{deAnnaPRF2017,zhaoPNAS2016}. The main topological differences between two-dimensional and three-dimensional structures rely on the contact point between the nearest grains (in a two-dimensional system different grains do not touch) which represent hot-spots for shear and fluid stretching and the chaotic advection resulting from the fluid mechanics analogue of the baker's map. The chaotic advection could impact scalar mixing when diffusion is considered~\cite{lester_dentz_leborgne_2016}. However, here we focus on the dynamical coupling between finite size particles transported by simple advection (no diffusion) and the carrying flow itself which does not depend on the peculiar flow kinematics and, thus, we expect that the discussed results are independent of the system dimension. Thus, to be comparable with previous results \cite{deAnnaPRF2017} and since the numerical method adopted is computationally expensive we decided to consider a two-dimensional heterogeneous structure, as described below.
	
\begin{figure}
		\centering
		\includegraphics[width=1\linewidth]{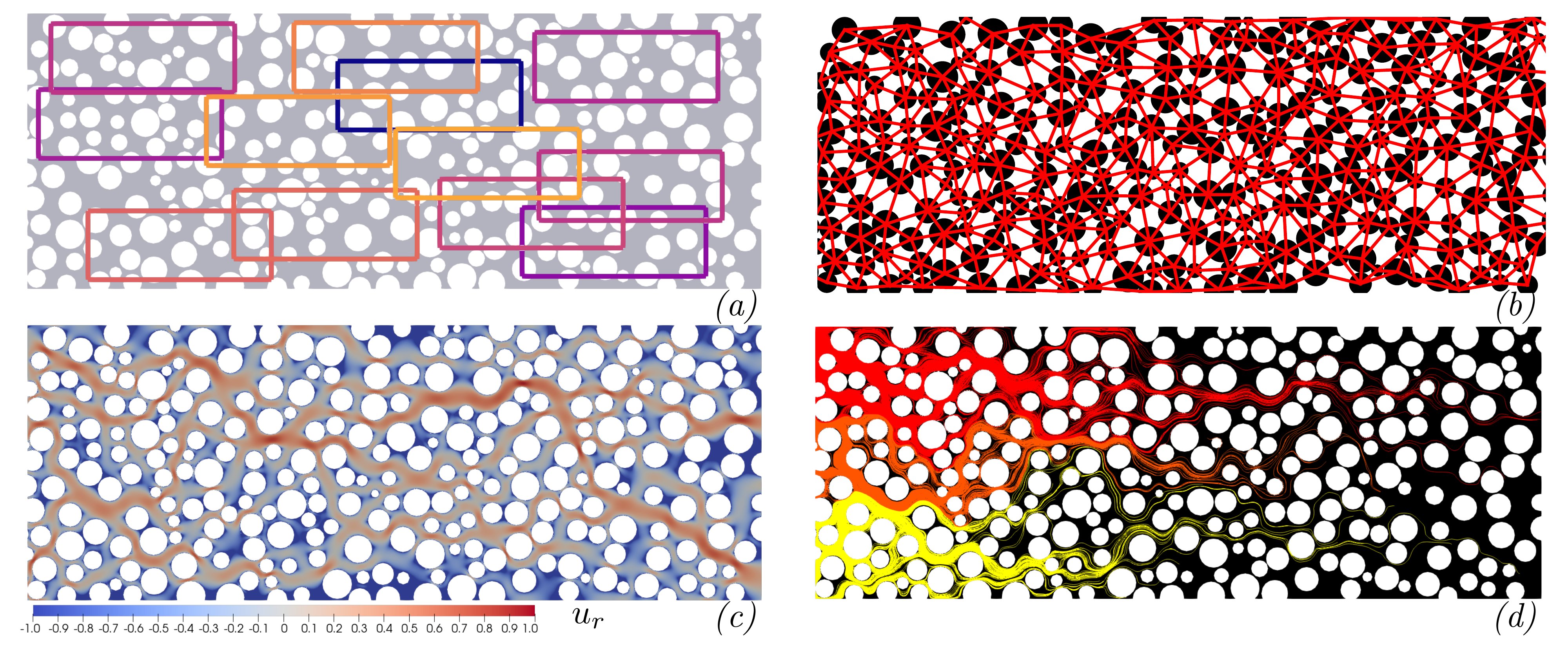}
		\caption{\emph{(a)} Geometry $\Omega$ and frames of the sub-geometries $\Omega_i$ considered for Stokes flow and fluid particle dynamics, respectively. The frames are color coded with respect to the velocity magnitude averaged over the sub-geometry, see Figure~\ref{fig2}.  \emph{(b)} Delaunay triangulation of $\Omega$, where edges are marked in red. Each edge corresponds to $d$. \emph{(c)} Rescaled velocity magnitude $u_r = \log_{10} (u/\langle u\rangle);$ $u = \lVert \u\rVert$, from eqs.~\eqref{eq1a}-\eqref{eq1b}. \emph{(d)} Streamlines of point particles initialized on the left side of $\Omega$. The different colors are for visualization purposes. They correspond to three equal size subdomains of the left boundary.}
		\label{fig0}
	\end{figure}

	To study the impact of suspended and finite-size particles on flow and transport we consider a two-dimensional porous domain $\Omega$ modeled by a rectangle $[x_0,x_1]\times[y_0,y_1]$ with cutout non-overlapping disks of random position and diameter, see Figure \ref{fig0}(a). This disordered arrangement of disks is characterized by a Delaunay triangulation of the disk centers to identify the nearest neighbors: each triangle defines a pore and each edge defines a throat, see Figure \ref{fig0}(b). The statistical properties of the structure are characterized in terms of pore throat size $\lambda = d - r_1 - r_2$ distribution: $d$ being the distance between the two neighboring disk centers and $r_1$ and $r_2$ being the respective radii. To consider a heterogeneous medium the domain structure is generated such that the PDF of the pore throat size roughly follows the power law distribution $p_\lambda(\lambda) \sim \lambda^{-\beta}$, with $\beta = 0.17$ as in \cite{deAnnaPRF2017}.

	Before we introduce the fluid particle dynamics model we consider Stokes dynamics. The results will be used to investigate potential differences and to validate the numerical approach against \cite{deAnnaPRF2017}.
	
	\subsection{Stokes dynamics}
	\label{stokes}
	Within this geometry we consider the Stokes equations 
	\begin{align}
		\label{eq1a}
		- \nu \nabla^2 \u + \nabla p &= \mathbf{0}, \\
		\label{eq1b}
		\nabla\cdot \u &= 0,
	\end{align}
	where $\u=\u(\mathbf{x})$ is the fluid velocity, $p=p(\mathbf{x})$ the fluid pressure, $\nu$ the fluid viscosity and $\mathbf{x} \in\Omega\subset\mathbb{R}^2$. Boundary conditions at $\partial \Omega$ are specified at the rectangle edges and the grain walls (perimeter of the disks). We impose $\u = \u_{\text{inflow}}$ on the left side of the rectangle $\partial \Omega_L$, $\nabla \u\cdot \mathbf{n} = \mathbf{0}$ on the right side of the rectangle $\partial\Omega_R$, $\u\cdot\mathbf{n} = \mathbf{0}$ on top $\partial\Omega_T$ and bottom $\partial\Omega_B$ rectangle boundaries and $\u = \mathbf{0}$
	on the remaining boundaries $\partial \Omega_{\text{disk}}$. Thereby, $\mathbf{n}$ denotes the outward normal to $\partial \Omega$. We consider 
	\[
	\u_{\text{inflow}}(x_0,y) = \left[\frac{1}{2} \left(1 + \tanh\left( \alpha - \left| y - \frac{y_1 + y_0}{2} \right|\right)\right), 0\right]^T,
	\]
	where $\alpha$ is chosen such that $\lVert\u_{\text{inflow}}(x_0,y_0)\rVert = \lVert\u_{\text{inflow}}(x_0,y_1)\rVert \leq 10^{-16}$ in order to numerically satisfy the compatibility condition with the top and bottom boundary conditions. The Dirichlet boundary condition imposed on $\partial\Omega_{L}$ differs from more commonly used pressure boundary conditions or a body force acting on $\partial\Omega_L$ and is chosen to be comparable with previous studies in \cite{deAnnaPRF2017}. For all of the aforementioned inflow modeling, the magnitude of the velocity field develops as a heterogeneous field characterized by channels with variable velocity and stagnation zones, which is what we want to reproduce.
	
	The problem is discretized in space with finite elements. To implement no-slip boundary conditions at solid interfaces and inflow conditions, we consider weak solutions of eqs. \eqref{eq1a} and \eqref{eq1b} with $\u\in \mathbf{V} := \{ \mathbf{v}\in H^1(\Omega)^2 : \mathbf{v} = \mathbf{0} \text{ at } \partial\Omega_{\text{disk}}, \, \mathbf{v} = \u_{\text{inflow}} \text{ at } \partial\Omega_L,\, \nabla\mathbf{u}\cdot\mathbf{n} = \mathbf{0} \text{ at } \partial\Omega_R\}$ and $p\in L^2(\Omega)$, such that 
	\begin{align}
		\label{eq1c}
		\into \nabla \v :\nu(\nabla\u + (\nabla\u)^T) - p\nabla\cdot\v\,\mathrm{d}\mathbf{x} &=  \mathbf{0}, \\
		\label{eq1d}
		\into q\nabla\cdot\u\,\mathrm{d}\mathbf{x}&=0,
	\end{align}
	for every $\v \in H^1_{\mathbf{0}}(\Omega)^2$ and  $q\in L^2(\Omega)$. We partition the domain $\Omega$ by a conforming triangulation $\mathcal{T}_h$. Then, the continuous spaces $\mathbf{V}\times L^2$ are approximated by the Taylor-Hood space $\mathbb{T}_h$  defined as \[
	\mathbb{T}_h = (\mathbb{V}_{2,g_0}\times \mathbb{V}_{2,g_1}) \times \mathbb{V}_1,\quad \!\!\! \mathbb{V}_{m,g} = \{v\in \mathbb{V}_m: \Tr_{\partial\Omega^\prime} v = g\},\quad \!\!\! \mathbb{V}_m =  \{p\in C^{0}(\Omega) : p|_{T} \in \mathbb{P}_m(T), \forall T\in\mathcal{T}_h\},
	\]
	where $\mathbb{P}_m$
	is the space of polynomials of order at most $m$ and $\partial\Omega^\prime = \partial\Omega \setminus \partial \Omega_R$. The velocity $\u$ and pressure $p$ are approximated by functions from $\mathbb{T}_h$, i.e. $m=2$ piece-wise quadratic for $\u$ and $m=1$ piece-wise linear for $p$. In the simulations we consider $\nu = 10^2$. We remark that because eqs.~\eqref{eq1a},\eqref{eq1b} are linear,  $\nu$ corresponds only to a rescale of the pressure $p$ and does not modify the velocity field $\u$. For further details about parameter setting see Section \ref{sec:parameters}.
	
	The resulting velocity field $\u$ is heterogeneous and organized into channels of high and fluctuating velocity and zones of stagnation. Figure~\ref{fig0}(c) shows the logarithm of $u_r= u/\langle u\rangle$ with $u = \lVert \u\rVert$ and $\langle \cdot \rangle$ the average. The color-map is such that blue (red) regions are associated with velocity magnitude below (above) its average value (gray). In order to visualize the transport properties within the medium we track the displacement of point particles along the streamlines, as shown in Figure~\ref{fig0}(d) where for each trajectory the color denotes its own initial vertical location, separated in three regions. 
	\begin{figure}
		\centering
		\includegraphics[width=1\linewidth]{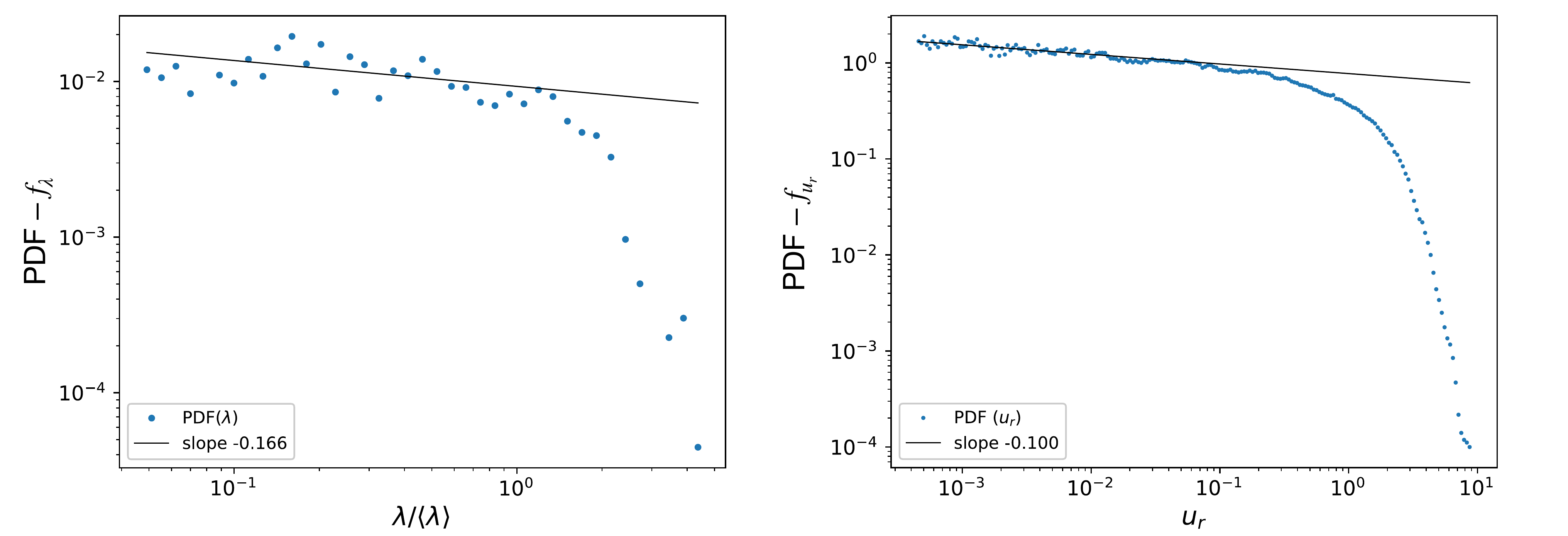}
		\caption{\emph{Left}. Double logarithmic plot of the probability density function (PDF) of the pore throat size distribution $f_{\lambda}$. The black line is a least square fit of the data. The theoretical distribution is $\lambda^{-\beta}$, $\beta = 0.166$, which match the fitted exponent. \emph{Right}. Double logarithmic plot of the probability density function of the velocities distribution $f_{u_r}$. The black line is a fit of the data for small velocities, i.e. $u_r < 10^{-2}$. The fitted line has slope $-0.100\pm 0.022$ in agreement with the prediction $f_{u_r}\sim u_r^{-\beta/2}$.}
		\label{fig1}
	\end{figure}
	Figure~\ref{fig1} shows the pore throat size distribution (left) and the distribution of the low velocities (right). The results reproduce a related power law distribution in the low range of pore throat size $f_{\lambda}\sim\lambda^{-\beta}$ and low velocities $f_{u_{r}}\sim u_r^{-\beta/2}$, with $\beta = 0.166$. The exponents are within the 95\% confidence interval of the analytic theory of \citep{deAnnaPRF2017}. 
	
	\subsection{Fluid particle dynamics}
	\label{method}
	
	\subsubsection{Mathematical model}
\begin{figure}
		\centering
		\includegraphics[width=1\linewidth]{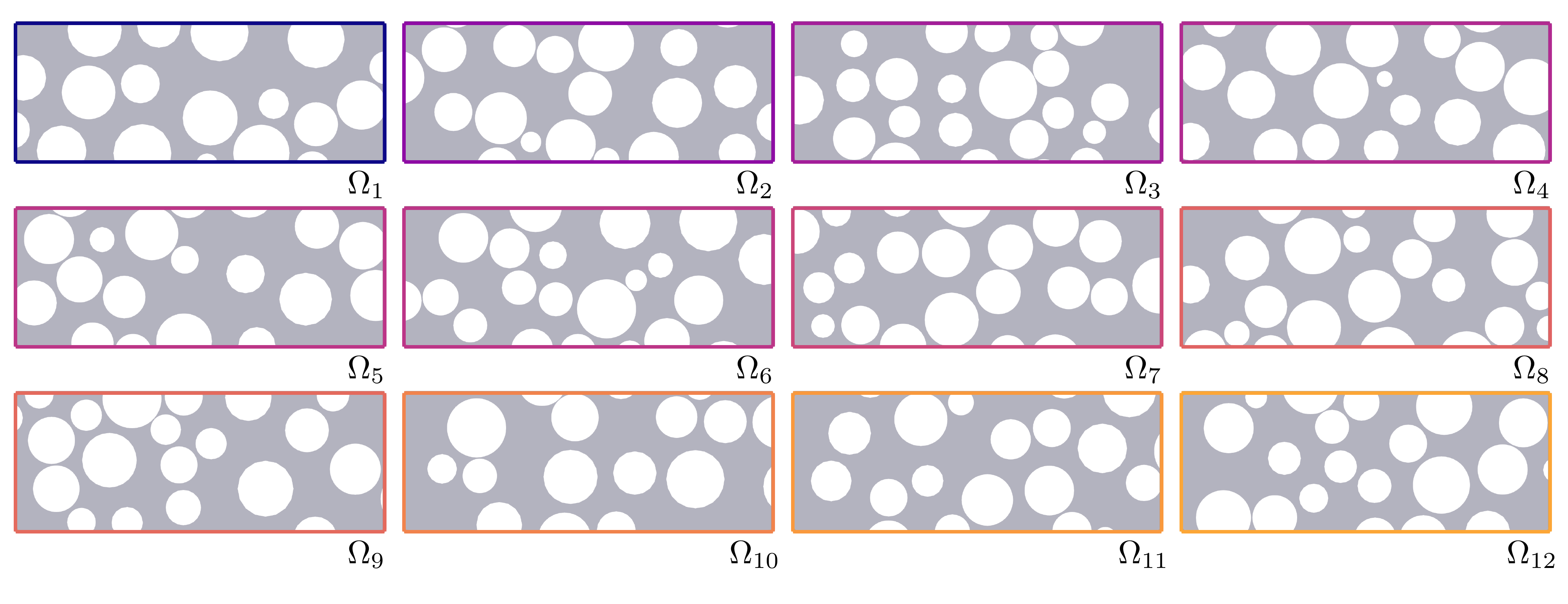}
		\caption{Different geometries $\Omega_i$, $i=1,\dots,12$ obtained from $\Omega$ in Figure~\ref{fig0}(a) ordered with respect to the average value of the Stokes flow $\langle u\rangle$ over the domain. The geometries consider different throat sizes but are too small to define a reasonable throat size distribution. }
		\label{fig2}
	\end{figure}

	To develop a model that takes into account the fluid-particle dynamics we follow the method propose by~\cite{tanaka00}. We consider the Navier-Stokes equations
	\begin{align}
		\label{eq2a}
		\rho (\u_t + (\u\cdot \nabla) \u) - \nabla\cdot(\nu(\nabla\u + \nabla\u^T)) + \nabla p &= \mathbf{f}, \\
		\label{eq2b}
		\nabla\cdot \u &= 0,
	\end{align}
	where $\u=\u(t,\mathbf{x})$ is the fluid velocity, $p=p(t,\mathbf{x})$ the fluid pressure, $\rho=const$ the fluid density, $\nu = \nu(t,\mathbf{x})$ the viscosity,  $\mathbf{f}=\mathbf{f}(t,\mathbf{x})$ the external volume forces and $(t,\mathbf{x}) \in (0,T]\times\Omega$. The external boundary conditions are the same as in the previous Section~\ref{stokes} with $\Omega$ replaced by sampled subdomains $\Omega_i$ to reduce the computational cost, see Figure \ref{fig0}(a) and Figure \ref{fig2}. 
	
	The fluid-particle dynamics approach describes particles as high viscosity regions within the fluid. This implies that the momentum transfer within the particle is much faster than through the fluid outside and it thus behaves like a solid particle. In order to deal with the viscosity contrast the particle is described by a diffuse domain approach. Let $\mathbf{x}_j = \mathbf{x}_j(t)$ be the position of the center of mass of particle $j$ of radius $a$. Then, the particle is represented through the concentration field
	\begin{equation}
		\phi_j(\mathbf{x}) = \frac{1}{2}\left(1 + \tanh\left(\frac{a - |\mathbf{x}-\mathbf{x}_j|}{\xi}\right)\right),\quad \mathbf{x}\in\Omega_i,
	\end{equation}
	where $\xi$ is the width of the diffuse interface. For $\xi \to 0$ the concentration field $\phi_j(\mathbf{x})$ approaches a characteristic function to represent the particle. Let $\nu_F$ and $\nu_P$ with $\nu_F \ll \nu_P$ be the fluid and particle viscosity, respectively. Then, the viscosity field $\nu$ is defined as
	\begin{equation}
		\nu(\mathbf{x}) = \nu_F + \sum_{j=1}^{N} (\nu_P - \nu_F)\phi_j(\mathbf{x}),\quad \mathbf{x}\in\Omega_i,
	\end{equation}
	where $N$ is the number of particles. In the limit case $\nu_P/\nu_F\to\infty$ and $\xi\to0$ we approach solid particle dynamics, see~\cite{tanaka00}. In this way, interactions between fluid and particles are encoded in the viscosity field. Particle-particle interactions are encoded in the force term $\mathbf{f}$. We consider the repulsive part of the Lennard-Jones potential, i.e.
	\begin{equation}
		V(\ell) = \varepsilon \left(\frac{\sigma}{\ell}\right)^{12}, 
	\end{equation}
	where $\varepsilon$ is the strength of the potential, $\sigma$ the interaction range and  $\ell$ the distance. The force acting on particle $j$ is given by 
	\begin{equation}
		\mathbf{f}_j = -\frac{\partial \sum_{k\neq j} V(|\mathbf{x}_j-\mathbf{x}_k|)}{\partial \mathbf{x}_j},\quad j=1,\dots N.
	\end{equation}
	Finally, the continuous force field, which enters the Navier-Stokes equations, is given by
	\begin{equation}
		\mathbf{f}(\mathbf{x}) = \sum_{j=1}^{N} \mathbf{f}_j\phi_j(\mathbf{x}).
	\end{equation}
	The position of particle $j$ is determined by
	\begin{equation}
		\label{eq3}
		\frac{\mathrm{d}\mathbf{x}_j(t)}{\mathrm{d}t} = \u_j(t)\quad \mbox{with} \quad \u_j(t) = \frac{\int_{\Omega}\u(t,\mathbf{x})\phi_j(\mathbf{x})\,\mathrm{d}\mathbf{x}}{\int_{\Omega}\phi_j(\mathbf{x})\,\mathrm{d}\mathbf{x}}.
	\end{equation}
	Eqs. \eqref{eq2a} - \eqref{eq3} together with the described boundary conditions and appropriate initial conditions define the model to solve. 
	
	The model contains several length scales (diffuse interface width $\xi$, particle radius $a$, average throat size $\langle\lambda\rangle$ and domain size $|\Omega_i|$), for which $\xi < a < \langle\lambda\rangle < |\Omega_i|$, and several time scales which range from fast particle-particle interactions to slow penetration times. All these different scales need to be resolved by the numerical method, which makes it computational expensive.
	
	In the limit of point particles (particle radius $a \to 0$) also the interaction range $\sigma$, which is related to $a$, goes to zero and the fluid-particle and particle-particle interactions vanish. In this limit eqs. \eqref{eq2a} - \eqref{eq3} can be approximated by the Stokes equations \eqref{eq1a} and \eqref{eq1b}. 
	
	The fluid particle dynamics method has the advantage of removing solid-fluid boundaries. This simplifies remarkably its implementation. A critical aspect of the approach employed in this work is that the artificial viscosity contrast $(\nu_F-\nu_P)\phi_j$ requires a small time-stepping in the discretization of the evolution equations.

	\subsubsection{Time discretization}
	As the finite size of the particles dynamically modifies the local flow field, the governing equation~\eqref{eq2a} explicitly depends on time $t$. To solve the flow evolution through time, we discretize time into $K$ steps of duration $\tau = T/K$ so that $t_k = k\tau \, \in [0,T]$, $k=0,\dots, K$. We consider a semi-implicit Euler method and an operator splitting approach to systematically solve first for the fluid flow and then for the particle positions integrating their motion equations~\eqref{eq3}. Given the value of $\u^{k}=\u(t_k,\mathbf{x})$, $\nu^k = \nu(t_k,\mathbf{x})$ and $\mathbf{f}^k = \mathbf{f}(t_k,\mathbf{x})$, the velocity field and pressure at the next time step, $\u^{k+1}$ and $p^{k+1}$, respectively, are approximated by
	
	\begin{align}
		\label{eq4a}
		\rho\left(\frac{1}{\tau}\u^{k+1} + (\u^{k}\cdot\nabla)\u^{k+1} + (\u^{k+1}\cdot\nabla)\u^{k}\right) - \nabla\cdot(\nu^k(\nabla\u^{k+1} +& (\nabla\u^{k+1})^T)) + \nabla p^{k+1} \nonumber \\
		& = \mathbf{f}^k + \rho\left(\frac{1}{\tau}\u^k + (\u^k\cdot\nabla)\u^k\right), \\\label{eq4b}
		\nabla\cdot\u^{k+1} &=0.
	\end{align}
	The velocity field $\u^{k+1}$ is used to update the particles position at the next time step by solving eq.~\eqref{eq3} with the explicit Euler method
	\begin{equation}
		\label{eq5}
		\mathbf{x}_j^{k+1} = \mathbf{x}_j^k + \tau \u_j^{k+1}. 
	\end{equation}

	\subsubsection{Space discretization} 
	The spatial discretization is accomplished by finite elements. We consider weak solutions of eqs. \eqref{eq4a} and \eqref{eq4b} with 
	$\u\in \mathbf{V}$ and $p\in L^2$, and $\Omega$ replaced by $\Omega_i$, such that 
	\begin{align}
		\label{eq6a}
		\intoi \rho\left( \frac{1}{\tau}\u^{k+1} + (\u^{k}\cdot\nabla)\u^{k+1} + (\u^{k+1}\cdot\nabla)\u^{k}\right) \v +&  \nabla \v :\nu^k(\nabla\u^{k+1} + (\nabla\u^{k+1})^T) - p^{k+1}\nabla\cdot\v\,\mathrm{d}\mathbf{x} \nonumber \\ &=  \intoi \left(\mathbf{f}^k + \rho\left(\frac{1}{\tau}\u^k + (\u^k\cdot\nabla)\u^k\right)\right)\v\,\mathrm{d}\mathbf{x}, \\
		\label{eq6b}
		\intoi q\nabla\cdot\u^{k+1}\mathrm{d}\mathbf{x}&=0,
	\end{align}
	for every 
	$\v\in H^1_0(\Omega_i)$ and $q\in L^2(\Omega_i)$. As for the Stokes equations we partition the domain $\Omega_i$ by a conforming triangulation $\mathcal{T}_h$. Then, the continuous spaces $\mathbf{V}\times L^2_0$ 
	are approximated by the Taylor-Hood space $\mathbb{T}_h$ with $\Omega$ replaced by $\Omega_i$. The velocity $\u^k$ and pressure $p^k$ are approximated by functions from $\mathbb{T}_h$, i.e. $m=2$ piece-wise quadratic for $\u^k$ and $m=1$ piece-wise linear for $p^k$.

	\subsubsection{Particle constraints and physical parameters} \label{sec:parameters}
	Expressing time in [\si{s}], space in [\si{cm}] and mass in [\si{g}], we used for the
	density the value of $\rho = 1$, as water, and for the fluid viscosity $\nu_F = 10^2$, which corresponds to a very viscous fluid.
	Thus, since the average pore size $\langle \lambda \rangle$ is about 50 and the average fluid velocity magnitude $\langle u \rangle$ is about 1, the characteristic Reynolds number of the system is $Re = \frac{ \langle \lambda \rangle \langle u \rangle \rho}{\nu_F} = 0.5$.
	Particles are represented as disks of high viscosity $\nu_P = 10^4$ with radius $a$ and center of mass $\mathbf{x}_j$. The values we have chosen for $\nu_F$ and $\nu_P$ are a compromise between computational feasibility and representing the physical properties of the system. The relation between $\nu_F$ and $\nu_P$ follows previous studies~\cite{tanaka00}.
	
	Particle positions $\mathbf{x}_j$ are updated through eq.~\eqref{eq5}. The new position results from averaging the velocity field $\u^{k+1}$ over the high viscosity region associated with the particle. As a consequence the distance between $\mathbf{x}_j$ and $\partial\Omega_i$ could be less than $a$ at the new time instance. To avoid this we impose a constraint on eq.~\eqref{eq5}. As soon as the distance between $\partial\Omega_i$ and $\mathbf{x}_j$ is less or equal to $a$, we adjust the velocity $\u^k_j$ by taking only the tangential component with respect to the boundary $\partial\Omega_i$.

	The fluid and particle densities have been matched and set to $\rho = 1$. This allows us to neglect any gravitational effect. We consider the potential strength $\varepsilon = \nu_P$ 
	and the interaction range $\sigma = 2 a$ in order to ensure effective repulsion. In order to deal with the singularity of the Lennard-Jones potential $V(x)$ at $x=0$ we consider $V(x) = V(2a)$ for $x < 2a$. We further truncate the potential such that $V(x) = 0$ for $x > (5/2a)^2$.
	
	The domain partition $\mathcal{T}_h$ is chosen in such a way that its elements remain as coarse as possible but are small enough in order to resolve the flow field. To ensure that the mesh size for fluid particle dynamics simulations is significantly smaller than $\xi$, we use adaptive refinement at the diffused particle interfaces. We chose $\xi = a/4$, where $a$ is the particle radius. The time step size $\tau$ has been chosen small enough so that the particle displacement never overcomes the particle radius, i.e. we impose $\tau \u_j^{k+1} < a$. Considering that at every time step the whole velocity field is recomputed and particles are advected, we have chosen $\tau = 0.5$. A similar finite element implementation of the fluid particle method, with similar parameters but simpler geometries can be found in~\cite{PV_JCP_2015}.
	
	\subsection{Implementation of the numerical scheme}
	\label{software}
	The numerical simulations are performed using AMDiS \cite{Vey_CVS_2007,witkowski15}. In its current version~\cite{amdis2} 
	it is based on the DUNE~\cite{sander20, Bastian2021} framework \url{https://www.dune-project.org/}. Particular features are adaptive refinement and wrappers from linear system solvers as PETSc. Both features are extensively used. We choose to discretize the domains $\Omega$ and $\Omega_i$ with the grid manager ALUGrid, see~\cite{alugrid}, and use the solver library MUMPS~\cite{mumps1,mumps2} imported by PETSc as a linear solver, see~\cite{petsc-web-page}. However, also with these advanced tools the fluid particle dynamics model cannot be solved with reasonable effort on the full domain $\Omega$ and is therefore considered on sampled subdomains $\Omega_i$, shown in Figure~\ref{fig2}. 
	
	\section{Results}
	\label{results}
	
	Our goal is to show how particle confinement, defined by the particle radius to averaged pore throat size ratio $a / \langle \lambda \rangle$, affects the velocity field $\u$ and the particles transport itself. As the particles radius is comparable to the constriction the particles have to pass through, the pore-throat, the fluid velocity is expected to dynamically change and to re-route fluid through other porous paths free of solid particles. This pore-scale phenomenology is also expected to breakdown the assumptions classically used that the laminar flow, dominated by viscous forces, transporting suspended particles is stationary~\citep{Molnar2005,Dentz2011,deAnnaPRF2017}. 
	We perform several simulations by systematically varying the particle radius $a \in \{0,1,2,4\}$. In Table~\ref{tab1} we report the average pore throat size for all considered geometries $\Omega_i$. We thus impose confinement conditions $a/\langle \lambda\rangle$ of $\sim 2\%$ ($a = 1$), $\sim 5\%$ ($a = 2$) and $\sim 10\%$ ($a = 4$), corresponding to realistic scenarios for suspensions moving through soil systems (e.g. a bacterial aggregate can easily reach a diameter of $10$ \si{\mu m} moving through pores of about $100$ \si{\mu m}).
	
	\begin{table}[]
		\centering
		\begin{tabular}{cccccccccccccc}
			\toprule
			Geometry & $\Omega$ & $\Omega_1$ & $\Omega_2$ & $\Omega_3$ & $\Omega_4$ & $\Omega_5$ & $\Omega_6$ & $\Omega_7$ & $\Omega_8$ & $\Omega_9$ & $\Omega_{10}$ & $\Omega_{11}$ & $\Omega_{12}$ \\
			\midrule
			$\langle\lambda\rangle$ & 56.791 & 42.539 & 42.673 & 47.410 & 47.924 & 53.106 & 53.729 & 53.633 & 42.748 & 48.904 & 52.591 & 52.127 & 52.996 \\
			\bottomrule
		\end{tabular}
		\caption{Average throat size $\langle\lambda\rangle$ 
			for $\Omega$ and the sampled sub-geometries.}
		\label{tab1}
	\end{table}
	
	In order to compare our simulations with each other we introduce the rescaled time $t/t^*$, where $t^* = \langle\lambda_i\rangle/\langle u_{S,i}\rangle$ with $\lambda_i$ the throat size and 
	\begin{equation}
		\label{eq:averageU}
		\langle u_{S,i}\rangle = \frac{1}{\Omega_i}\int_{\Omega_i} \lVert \u_{S,i} \rVert\,\mathrm{d}\mathbf{x}
	\end{equation}
	the average velocity magnitude of the Stokes flow over the geometry $\Omega_i$, respectively. Then, we study the evolution of the system for $t\in[0, 1000]$, corresponding to a final time $1000/t^\ast \in [10,29]$.
	\begin{figure}
		\centering
		\includegraphics[width=1\linewidth]{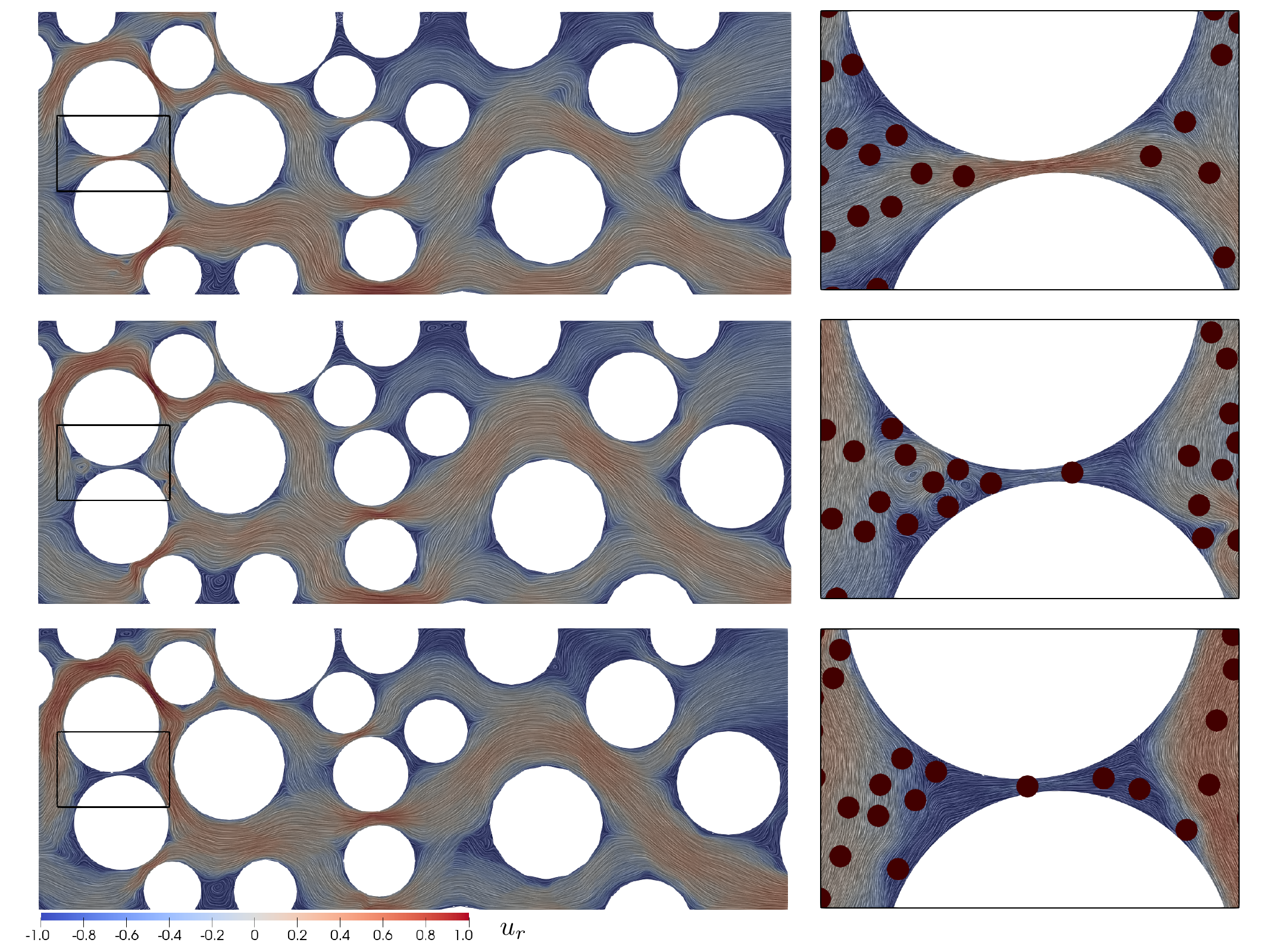}
		\caption{\emph{Left}. Snapshots of $u_r$ of the fluid particle dynamics model at different times $t/t^* = 6.42, 19.26, 28.90$ (t = 200, 600, 900, from top to bottom). The velocity field is visualized with the line integral contour technique (LIC) and color coded by the rescaled velocity using eq. \eqref{eq:ur}. \emph{Right}. Magnification of the region identified by the black box. In particular, particles are visualized as dark dots with radius $a$. On top we observe typical fluid flow through the channel. In the middle magnified particle-particle interactions are highlighted and the effect on fluid flow is visible, e.g. by vortexes on the left side of the channel. At the bottom a particle almost blocks the flow through the pore throat. These results correspond to $\Omega_9$ and confinement condition $a/\langle\lambda\rangle\sim 10\%$.}
		\label{fig3}
	\end{figure}
	Figure~\ref{fig3} shows snapshots of the rescaled velocity magnitude
	\begin{equation} \label{eq:ur}
		u^k_r:=\log_{10}( \lVert\u^k\rVert/\langle u_{S,i}\rangle), 
	\end{equation}
	for $\Omega_9$ at three time instances 
	and a confinement $a/\langle\lambda\rangle\sim 10\%$. The color-code is such that blue (red) zones are regions where the velocity magnitude is below (above) the average velocity magnitude (gray). As for the Stokes solution under no confinement ($a/ \langle \lambda \rangle = 0$), the velocity field exhibits high velocity channels and stagnation zones. However, this heterogeneity now changes over time. In particular, the magnified pictures show that fluid-particle and particle-particle interactions significantly influence the velocity field. Laminar vortex structures appear for $t/t^* = 19.26$ ($t = 600$) left of the pore-throat. They result from fluid re-routing towards other, more permeable paths. The fluid-particle and particle-particle interactions can also lead to a dramatic flow reduction almost stopping fluid motion in some pore-throats, as seen at $t/t^* = 28.90$ ($t = 900$), where this situation results from a particle almost blocking the flow through the pore-throat. 
	
	\subsection{Velocity field}
	
\begin{figure}
		\centering
		\includegraphics[width=1\linewidth]{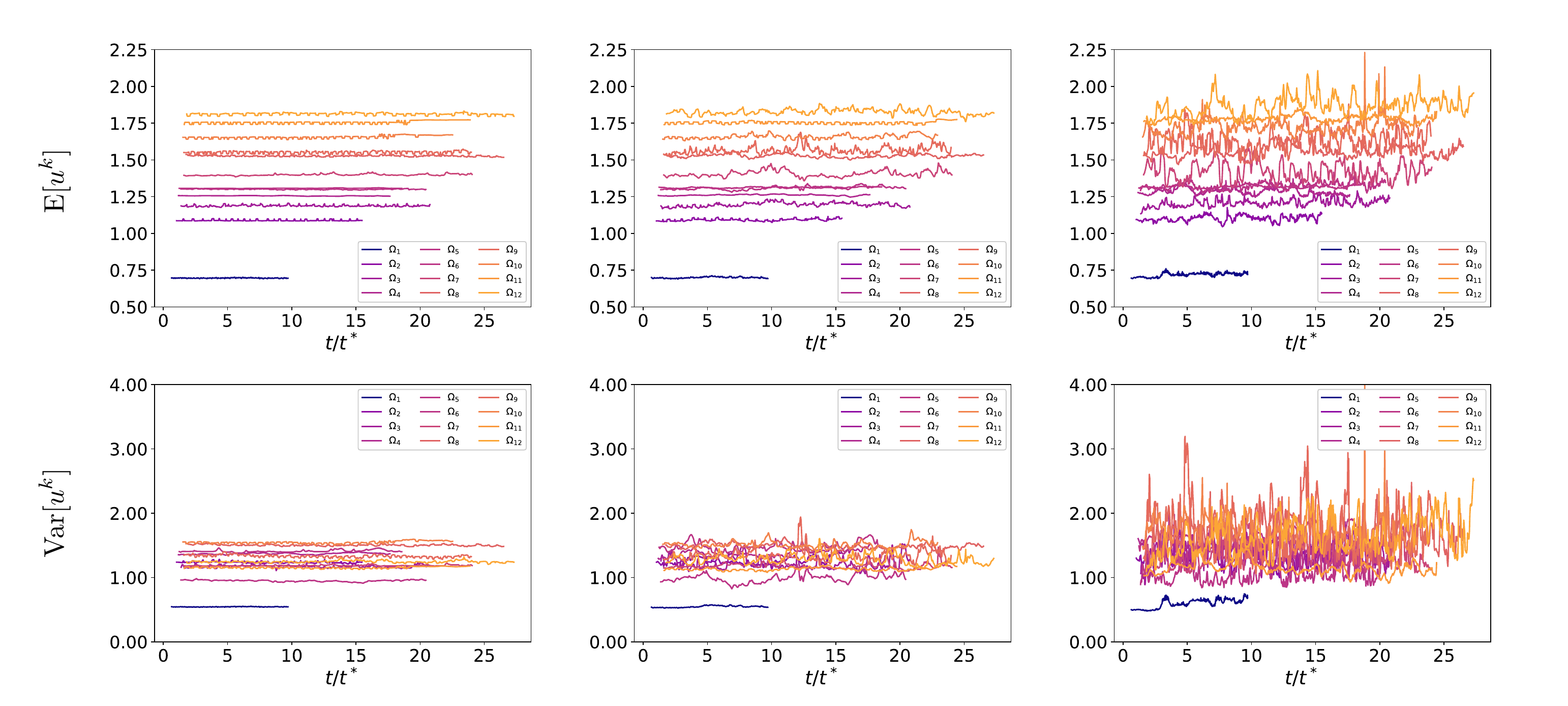}
		\caption{\emph{Top}. From left to right the mean of the PDF of $u^k$ for the different geometries $\Omega_i$ for the confinement $a/\langle\lambda\rangle$ $\sim 2\%$ (\emph{left}), $\sim 5\%$ (\emph{mid}), $\sim 10\%$ (\emph{right}). On the x-axis is the rescaled time $t/t^*$, $t^* = \langle\lambda\rangle/\langle u_{S,i}\rangle$, where $\langle u_{S,i}\rangle$ is the average velocity field of the static Stokesian flow corresponding to the domain $\Omega_i$. \emph{Bottom}. The variance of the PDF of $u^k$. It is evident that for small confinement ($\sim 2\%$) the mean and variance are almost constant, while for larger confinements ($\sim 5\%$ and $\sim 10\%$) the values oscillate with increasing intensity. }
		\label{fig4}
		
	\end{figure}

	To analyze the temporal variability of the velocity field $\u$, we consider the spatial mean of $u^k = \lVert\u^k\rVert$ and plot it against the rescaled time $t/t^*$. The PDFs are constructed by considering a fixed bin range $[b_l, b_r]$ for all $\Omega_i$, where $b_r = \max_i u_{S,i}$ and $b_l = b_r/10^4$.
	The bins are logarithmically distributed in $[b_l, b_r]$. The mean $\textrm{E}[u^k]$  and variance $\textrm{Var}[u^k]$ of the PDFs are shown in Figure~\ref{fig4}.  These results suggest that the mean velocity oscillates mainly around a constant value. However, the variance strongly depends on the confinement and increases significantly with increasing confinement $a / \langle \lambda \rangle$. 

\begin{figure}
		\centering
		\includegraphics[width=1\linewidth]{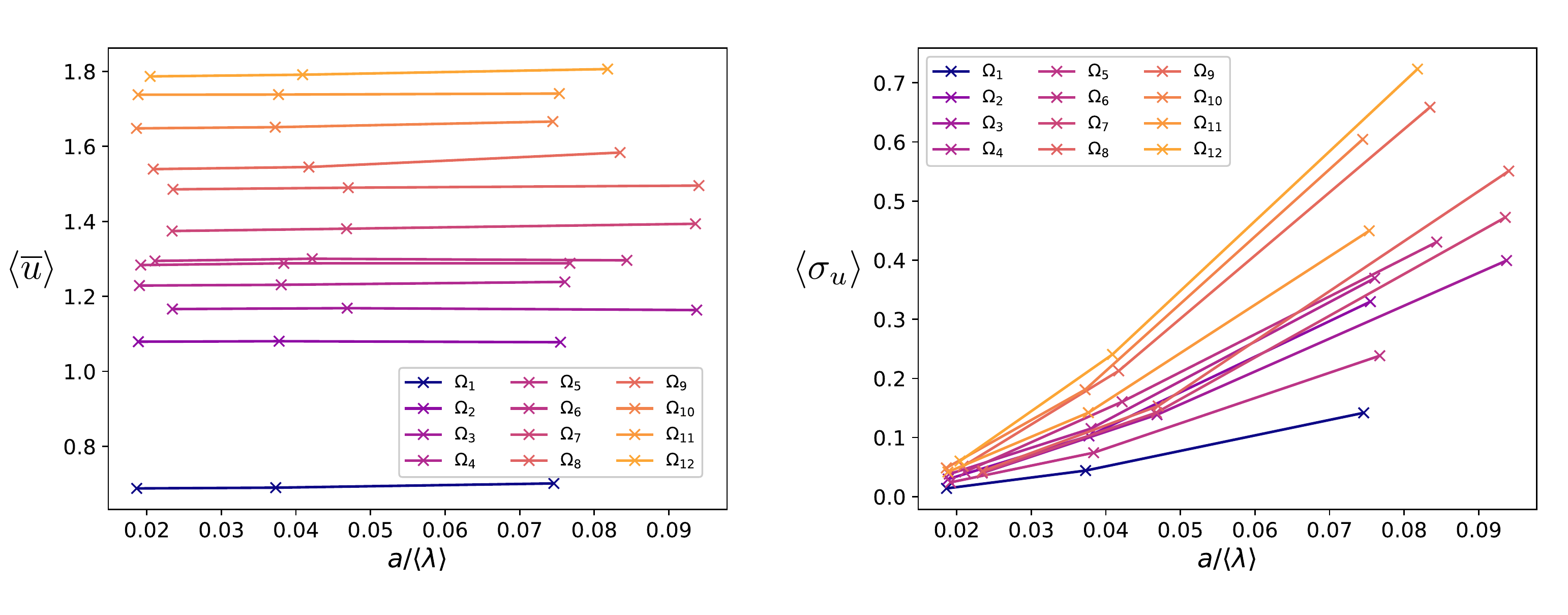}
		\caption{For the geometries $\Omega_i$ in Figure~\ref{fig2} the quantities $\langle\bar{\u}\rangle$ (\emph{left}) and $\langle\sigma_{\u}\rangle$ (\emph{right}) are displayed. The numbering and color coding of the subdomains is ordered with respect the the averaged velocity. }
		\label{fig5}
	\end{figure}

\begin{figure}
		\centering
		\includegraphics[width=1\linewidth]{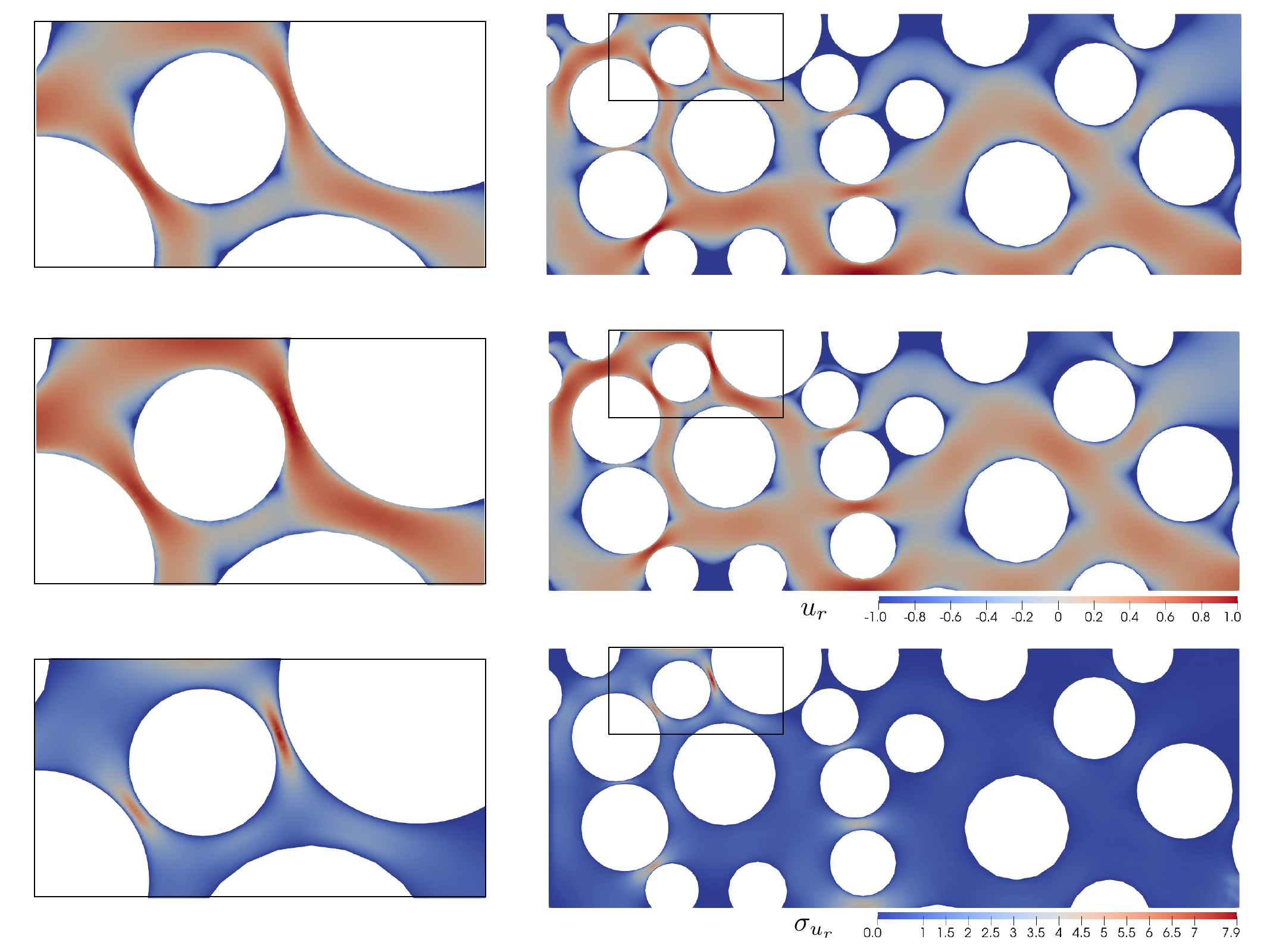}
		\caption{\emph{Top.} Color plot in logarithmic scale of the rescaled velocity $u_r$ for the Stokes flow with a magnification of the region identified by the black box. \emph{Middle.} Average flow $u_r$ obtained by the fluid particle dynamics approach with particles of radius $a=4$ (particle confinement $\sim 10\%$). The resulting velocity fields are similar. This is justified by the fact that high velocity channels of the dynamic flow flip repetitively, resulting, on average, in similar flow profiles as the Stokes flow. \emph{Bottom.} Color plot of $\sigma_{\mathbf{u}}$. Notice that the velocity field has mostly changed within the high velocity channels. See also the magnified region. }
		\label{fig6}
	\end{figure}

	To confirm this we consider the time-average velocity field $\bar{\u}$ and its variance $\sigma_{\u}$, as
	\begin{equation}
		\bar{\u} = \frac{1}{K}\sum_{k=1}^{K}\u^k,\qquad \sigma_{\u} = \sqrt{\frac{1}{K}\sum_{k=1}^{K} \lVert\u^k -\bar{\u}\rVert^2}.
		\label{eq9}
	\end{equation}
	Figure~\ref{fig5} shows both quantities as functions of the confinement $a/ \langle \lambda \rangle$ for different subdomains $\Omega_i$. The space-time averaged velocities are constant with respect to the confinement, but the velocity field changes over time and these changes increase with confinement. While the actual values depend on the considered subdomain $\Omega_i$ and with it the actual pore structure, the qualitative behavior remains the same for all $\Omega_i$. 
	Both quantities are also shown in Figure~\ref{fig6} for the geometry $\Omega_9$ and particle confinement $a / \langle \lambda \rangle \sim 10\%$. For comparison, we consider also the stationary Stokes equations with point-like particles for the same geometry and parameter setting. The time average velocity field corresponds to the solution of the Stokes equations, while its variance strongly differs mostly in a few pores characterized by high velocity. This reflects the microscopic effect of particle-fluid interactions under confinements. When particles approach a constriction they interact with the viscous fluid modifying the local velocity field and re-routing the flow (and the particles themselves) towards another permeable path nearby. Thus, the resulting velocity field variability is stronger in channels of higher permeability and velocity.
	
	\subsection{Transport properties}
	
\begin{figure}
		\centering
		\includegraphics[width=1\linewidth]{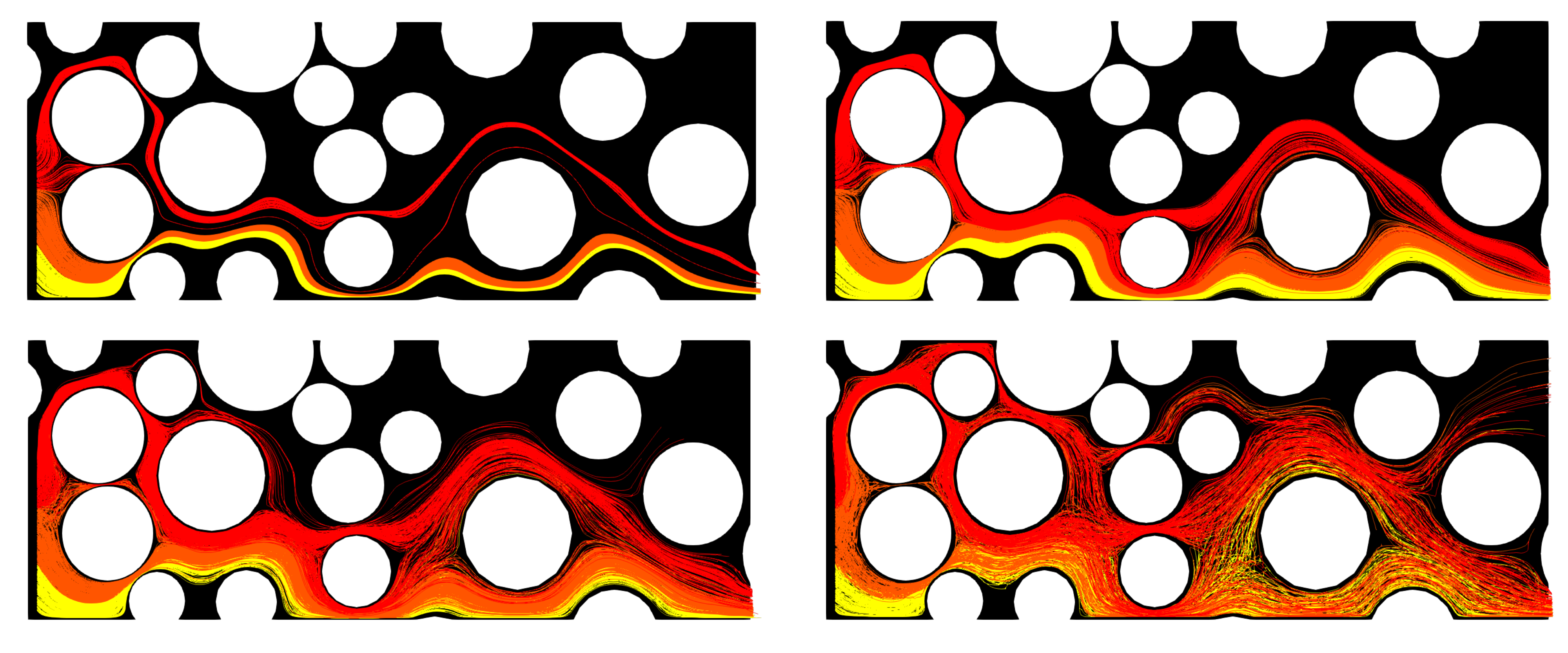}
		\caption{Particle streamlines for the geometry $\Omega_9$ and different particle radius $a$: point particles ($a=0$) (top-left), $a = 1$ (top-right), (bottom-left) $a=2$, (bottom-right) $a=4$. The different colors are for visualization purposes. They correspond to three equal parts of the left boundary.}
		\label{fig7}
	\end{figure}

	We track the suspension particles over time to illustrate their effect on the velocity field as well as their own transport properties.
	Figure~\ref{fig7} visualizes particle streamlines at $1000/t^*$ for the geometry $\Omega_9$: for point-like particles, and particles of radius $a=1, 2, 4$.  The starting locations of particles on the left boundary are the same for all simulations and are flux-weighted~\citep{deAnnaPRF2017}. However, the trajectories develop in a very different manner depending on the pore structure. Due to the finite size of particles and their interaction with the fluid, trajectories tend to spread over the computational domain. The color coding is as in Figure~\ref{fig0}(d). While for point-like particles the divergence free velocity field implies that trajectories never cross, for finite size particles this is no longer true. See the mixing colors in Figure~\ref{fig7}. This effect increases with the suspended particle radius. The trajectory spreading clearly invades low velocity regions which are poorly invaded by point particles. 
	
\begin{figure}
		\centering
		\includegraphics[width=1\linewidth]{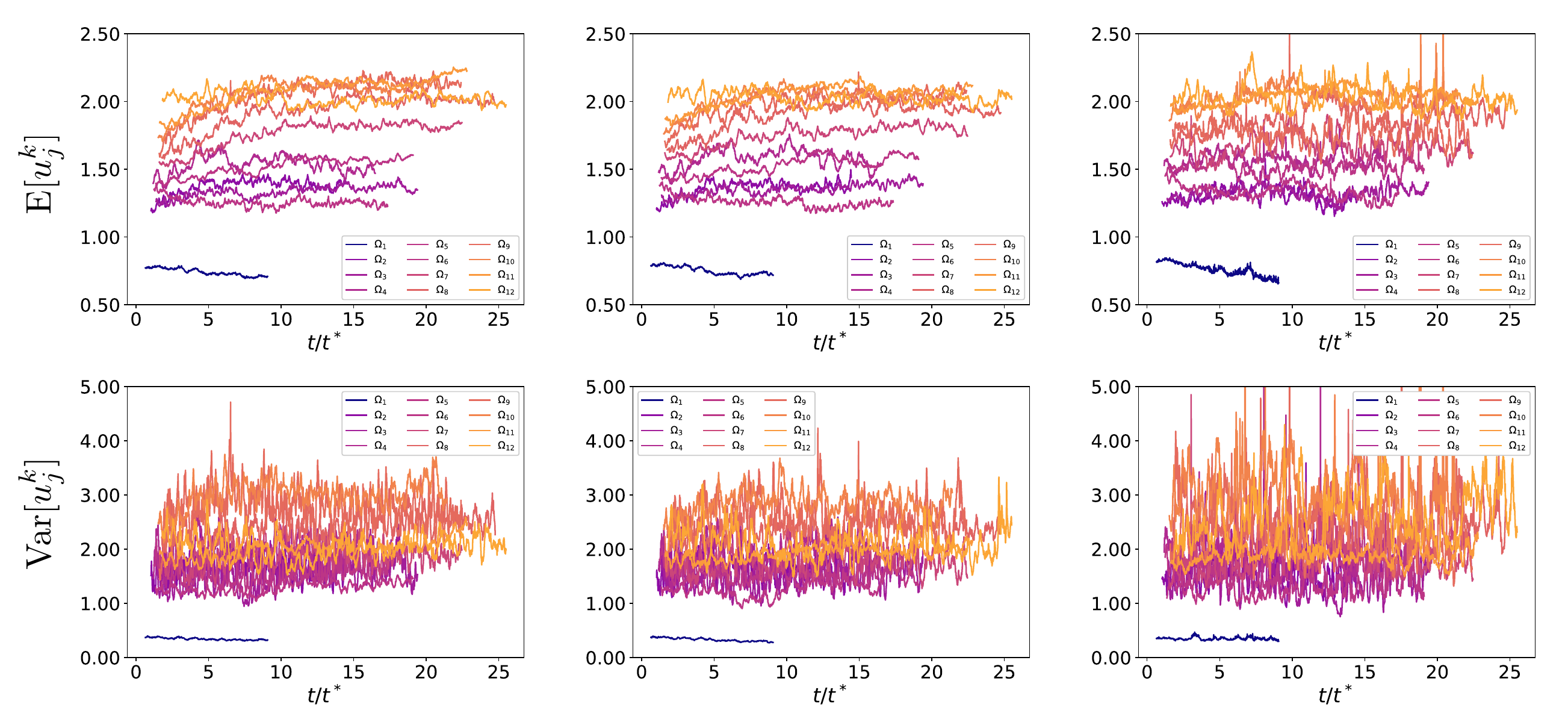}
		\caption{At each time step the PDF of the particle velocity magnitude is computed. On the x-axis is the rescaled time $t/t^*$, where $t^* = \langle \lambda\rangle/\langle u\rangle$ with $u$ magnitude velocity of the Stokesian flow. Notice that we have different time scales for different geometries since the throat size average varies with $\Omega_i$. On the y-axis the  mean (\emph{top}) and variance (\emph{bottom}) of the PDF for different particle confinements $\sim 1\%$ (\emph{left}), $\sim 5\%$ (\emph{middle}) and $\sim 10\%$ (\emph{right}) are shown.}
		\label{fig8}
	\end{figure}

\begin{figure}
		\centering
		\includegraphics[width=1\linewidth]{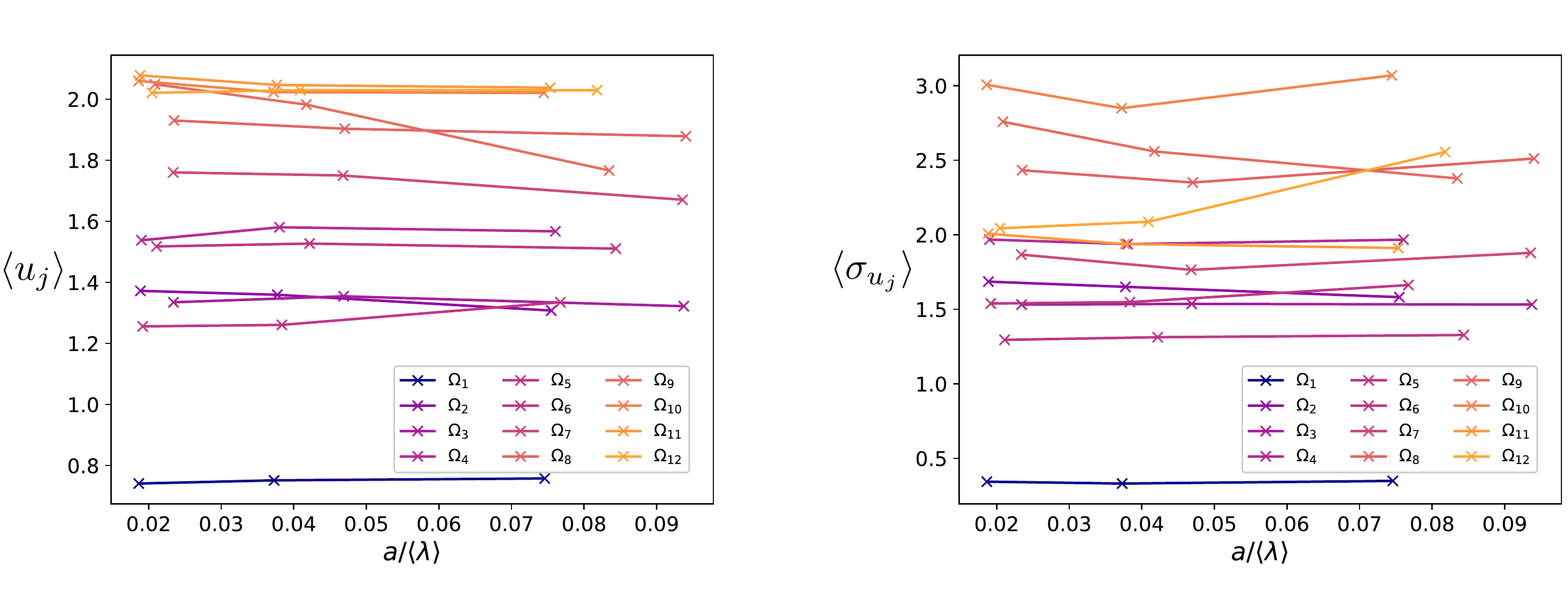}
		\caption{The space-time average of the particle velocities (\emph{left}) and their variance (\emph{right}) is shown for all sub-geometries $\Omega_i$ as a function of the particle confinement $a/\langle \lambda \rangle$.}
		\label{fig9}
	\end{figure}

	We next conduct the same analysis as for the velocity field but now restricted to particles. In particular, for each geometry $\Omega_i$ we consider the PDF of the particle velocities magnitude $u_j^k =\lVert \u_j(t^k)\rVert$ (see~\ref{eq5}) and compute the mean and variance at every time step $t^k$ averaged over all trajectories, shown in Figure~\ref{fig8}. The mean particle velocity magnitude $\mathrm{E}[u^k_j]$, Figure~\ref{fig8} (\emph{top}), is larger than the corresponding mean of the velocity field $\mathrm{E}[u^k]$ in Figure~\ref{fig4} (\emph{top}). This can be explained as particles are most likely to be found in high velocity channels. The numerical simulations suggest that the mean particle velocity mainly oscillates around a constant value during the time evolution for $t/t^*$ sufficiently large. Differently from the variance of the velocity field magnitude $\mathrm{Var}[u^k]$ in Figure~\ref{fig4} (\emph{bottom}), the variance of the particles velocity magnitude $\mathrm{Var}[u^k_j]$ (Figure~\ref{fig8} (\emph{bottom})) does not increase systematically with particle confinement $a/\langle\lambda\rangle$. This indicates that particles remain in high velocity channels even if the flow changes. 
	In order to better highlight these effects we further average over time and plot the results in Figure~\ref{fig9}. The space-time averaged velocity $\langle u_j \rangle$ and its variance $\langle\sigma_{u_j}\rangle$ remains essentially constant with respect to particle confinement.
	
	
	To provide a quantitative measure of individual displacement, we compute the average particles velocity along the longitudinal, $x$, direction. Let $t^0_j$ and $t^*_j$ be the spawn and stop times of the individual particle $j$. The spawn time corresponds to the injection time of particle $j$ at the left side of the domain, and the stop time $t^*_j$ corresponds either to the end of the simulation time or the time at which the particle $j$ reaches the right side of the computational domain. Let $[x_j(t), y_j(t)]$ be the position of the particle $j$ at time $t$. Then, the average velocity along the $x$-direction is given by
	\[
	\bar{v} = \frac{1}{J}\sum_{j=1}^J v_j, \quad v_j = \frac{x_j(t^*_j) - x_j(t^0_j)}{t^*_j - t^0_j}.
	\]
	Table~\ref{tab0} gives $\bar{v}$ for the different geometries and radii. Notice that as particle confinement increases, the propagation velocity decreases for most of the considered geometries.

	\begin{table}[]
		\centering
		\begin{tabular}{c S[round-mode = places, round-precision=3] S[round-mode = places, round-precision=2] S[round-mode = places, round-precision=3] S[round-mode = places, round-precision=2] S[round-mode = places, round-precision=3] S[round-mode = places, round-precision=2] S[round-mode = places, round-precision=3] S[round-mode = places, round-precision=2] S[round-mode = places, round-precision=3] S[round-mode = places, round-precision=2] S[round-mode = places, round-precision=3] S[round-mode = places, round-precision=2]}
			\toprule
			{Radius} & \multicolumn{1}{c}{$\Omega_1$} & ~ & \multicolumn{1}{c}{$\Omega_2$} & ~ & \multicolumn{1}{c}{$\Omega_3$} & ~ & \multicolumn{1}{c}{$\Omega_4$} & ~ & \multicolumn{1}{c}{$\Omega_5$} & ~ & \multicolumn{1}{c}{$\Omega_6$} \\
			\midrule
			$0$ & 0.7906071596988675 & ~ & 1.54276553638329 & ~ & 1.2657443869720968 & ~ & 1.6975316907900715 & ~ & 1.905573392489862 & ~ & 1.4698532056582503 \\
			$1$ & 0.7213495626108003 & {(}0.08760052{)} & 1.500168091515123 & {(}0.02761109{)} & 1.1756218184400042 & {(}0.07120124{)} & 1.6872235072472899 & {(}0.00607245{)} & 1.7651761597640374 & {(}0.07367716{)} & 1.4605231967094003 & {(}0.00634758{)} \\
			$2$ & 0.7079611880624465 & {(}0.10453481{)} & 1.517819191873465 & {(}0.01616989{)} & 1.1918769000045217 & {(}0.05835893{)} & 1.7341340357478723 & {(}-0.0215621{)} & 1.7891522440634902 & {(}0.06109507{)} & 1.4450446072867347 & {(}0.01687828{)} \\ 
			$4$ & 0.6064515649217492 & {(}0.23292933{)} & 1.3837423335447983 & {(}0.10307671{)} & 1.1067125620969998 & {(}0.12564292{)} & 1.6884675469216002 & {(}0.0053396{)} & 1.887946101198029 & {(}0.00925039{)} & 1.303872851411751 & {(}0.11292308{)} \\
			\bottomrule
		\end{tabular}
		\vspace{2mm}
		
		\begin{tabular}{c S[round-mode = places, round-precision=3] S[round-mode = places, round-precision=2] S[round-mode = places, round-precision=3] S[round-mode = places, round-precision=2] S[round-mode = places, round-precision=3] S[round-mode = places, round-precision=2] S[round-mode = places, round-precision=3] S[round-mode = places, round-precision=2] S[round-mode = places, round-precision=3] S[round-mode = places, round-precision=2] S[round-mode = places, round-precision=3] S[round-mode = places, round-precision=2]}
			\toprule
			{Radius} & \multicolumn{1}{c}{$\Omega_7$} & ~ & \multicolumn{1}{c}{$\Omega_8$} & ~ & \multicolumn{1}{c}{$\Omega_9$} & ~ & \multicolumn{1}{c}{$\Omega_{10}$} & ~ & \multicolumn{1}{c}{$\Omega_{11}$} & ~ & \multicolumn{1}{c}{$\Omega_{12}$} \\
			\midrule
			$0$ & 1.832855034108713 & ~ & 1.8716197946468156 & ~ & 2.2477695314743884 & ~ & 2.3629207121435023 & ~ & 1.9564792890567035 & ~ & 1.8054641979118713 \\ 
			$1$ & 1.8468361517991825 & {(}-0.00762805{)} & 1.961274951056896 & {(}-0.04790244{)} & 2.2203308488447138 & {(}0.01220707{)} & 2.283732214992107 & {(}0.03351297{)} & 1.9431316491252366 & {(}0.00682228{)} & 1.776941447054544 & {(}0.01579802{)} \\ 
			$2$ & 1.7373948227911884 & {(}0.05208279{)} & 1.854612679764501 & {(}0.00908684{)} & 2.0209583512584373 & {(}0.100905{)} & 2.2343033215605876 & {(}0.05443153{)} & 1.9653440411508039 & {(}-0.00453097{)} & 1.7755826519776705 & {(}0.01655062{)} \\ 
			$4$ & 1.494538644317411 & {(}0.18458437{)} & 1.5884510430148324 & {(}0.15129609{)} & 1.4949981110652844 & {(}0.33489707{)} & 2.046028142185807 & {(}0.13411054{)} & 1.8134428772166753 & {(}0.07310909{)} & 1.7457076631682573 & {(}0.0330976{)} \\ 
			\bottomrule
		\end{tabular}
		\caption{
			Longitudinal ($x$-direction) average propagation velocity $\bar{v}$ of particles for different geometries $\Omega_i$ and radii $a$. Within brackets is reported the value $\chi = 1 - \bar{v}_a/\bar{v}_0$, where $\bar{v}, \bar{v}_0$ are the average velocities along the $x$-direction of particles with radius $a$ and $0$ (point particles), respectively. A positive value implies that particles with radius $a$ have propagated faster than the corresponding point particle and vice-versa for a negative value.}
		\label{tab0}
	\end{table}
	
	\section{Conclusions}
	\label{conclusions}
	
	We have simulated the transport of soft particles of finite size through 2D porous geometries systematically changing the confinement conditions, represented by the ratio between the particle radius and the average pore-throat size of the pore structure. The numerical approach employed is the fluid particle dynamics approach introduced in~\cite{tanaka00}. The numerical scheme allows us to simulate the Navier-Stokes equations with particle interactions without explicitly treating particle boundaries. We found that particle confinement affects the fluid velocity field $\u$ which in turn affects the particles transport. We show that these fluid-particle interactions significantly impact the overall transport. 
	
	Microscopically, as shown in Figure~\ref{fig7}, even under small confinement conditions, the fluid and the transported particles are dynamically re-routed towards more permeable paths in a dynamical way. This is expected to have a significant impact on transport-driven phenomena associated with the particles themselves or the solutes dissolved in the fluid. Among these phenomena we highlight i) the formation/dissipation or persistence of solute gradients, ii) mixing, iii) chemical reaction with the solid grains or iv) filtration. In particular, we expect the latter to be affected by this phenomenon. This is because classical filtration theories do not take into account flow variability and assume stationary conditions, which we show here to be disrupted by confinement. Moreover, this leads to the emergence of ephemeral laminar vortexes at pore throat entrances (as shown in Figure~\ref{fig3}), a qualitative property that could affect solute transport and their gradient dynamics.
	
	Macroscopically, table~\ref{tab0} shows the impact of confinement on the overall velocity experienced by transported particles along the longitudinal $x$-direction. More specifically, table~\ref{tab0} reports the value of the average propagation velocity $\bar{v}$ of particles in each geometry $\Omega_i$ and radius $a$. Then, the quantity $\chi = 1 - \bar{v}_a/\bar{v}_0$ is computed, which measures the variability of the overall velocity experienced by particles of size $a$ compared to their point-like counterparts. A positive value of this quantity implies that particles of radius $a$ have propagated faster than point-like particles and a negative value implies the opposite. Among the 12 cases investigated for the 4 confinement cases considered, most showed positive results.
	This means that confinement has the net effect of forcing particles to move significantly faster through the medium: $0.025$ is the average of $\chi$ over the 12 geometries for $a = 1$ and $0.12$ for $a = 4$ corresponding to weaker and stronger confinement. This is expected to impact the overall arrival times and breakthrough curves in larger porous systems.
	
	
	The complexity of the fluid-suspended particles interaction, flow hydrodynamics, the coupling between particles size, local fluid velocity and associated transport have been widely overlooked. Our results estimate the effect of these phenomena on macroscopic flow kinematics and transport which are relevant for mixing, reaction kinetics, filtration and bacterial transport that often are found as aggregates of variable size. We expect these results to be also of relevance in more complex scenarios where also morphological variability, in terms of grain shape \citep{Alhashmi2016,Xiong2016,Wu2019} plays critical roles, e.g. in groundwater contamination and remediation~\cite{Kahler2019}, enhanced hydrocarbon recovery~\cite{Kar2015}, transport through river sediments that create a closely packed pore network~\cite{Lei2022}, water filtration systems~\cite{Kosvintsev2002} and extra-cellular transport in brain tissue~\cite{Nicholson2017}. Also, for pores surrounded by a single grain that cannot host a net transfer of fluid, the so-called dead-end pores, which have been recently shown to host laminar flow vortexes and are able to trap fluid for long times~\cite{Bordoloi2022}, it would be interesting to estimate the impact of fluid-suspended particles interaction and flow hydrodynamics. \\
	
	{\bf Acknowledgments:} This work was funded by the EU H2020 program within FET-OPEN project NARCISO (Grant Agreement no. 828890). We further acknowledge computing resources at FZ Jülich under grant PFAMDIS and at ZIH under grant WIR.
	
	\newpage
	\bibliographystyle{unsrt}

\end{document}